# CHARACTERISTICS OF UNDULATOR-TYPE RADIATION EMITTED BY BUNCH OF CHARGED PARTICLES IN WAKEFIELD

*Anatoliy Opanasenko*

*National Science Center "Kharkov Institute of Physics & Technology", Kharkov, Ukraine*
e-mail: Opanasenko@kipt.kharkov.ua

We consider spectrum-angular characteristics of the undulator-type radiation emitted by a bunch of relativistic charged particles because of interacting with the nonsynchronous spatial harmonics of the transverse wakefields excited by this bunch as it moves in a periodic structure. The conditions acceptable for possible experimental verification when incoherent undulator-type radiation power exceeds power loss associated with exciting the wakefields are discussed.

PACS: 29.27, 41.60.A, 41.60.B, 41.60.C

## 1. INTRODUCTION

Studies of the undulator type radiation, emitted by a beam of relativistic charged particles undulating in the alternating wakefields induced by this beam as it moves in a periodic structure, is an interesting area of accelerator and FEL physics. Recently it has received certain development in the works [1-5]. Two new radiation mechanisms have been proposed for generating ultra-short wavelength light. In Refs. [1,2] the quasi-static image charge fields produced by sheet electron beam passing between two periodic grating surfaces were considered as wiggler type fields, and a new device called the image charge undulator was proposed. These image charge fields dominate when the bunch length much longer than the maximal wavelength $\lambda_{max}$ of the Cherenkov-type radiation, CR, emitted by an electron in the periodic structure. (The term CR covers different names of the radiation, emitted by a charged particle moving with constant velocity in periodic structures, such as Smith-Purcell, diffraction, quasi-cherenkov, resonant-transition, parametric-cherenkov radiation, etc. [6].)

When the bunch length shorter than $\lambda_{max}$, the fields of the coherent Cherenkov-type radiation become prevalent wakefields, WF. In this case, as it has been shown in Refs. [3-5], transverse components of the nonsynchronous spatial wakefield harmonics acting on particles can give rise to their undulating motion and consequently to generating the undulator-type radiation, UR. The theory of the fundamental mechanism of the radiation emitted by ultrarelativistic charged single-particle in its own wakefield induced in an infinitely long periodic structure was given in [4,5]. As follows from the theory, in the relatively long-wave spectral region, where diffraction of generated waves is essential, the radiation manifests itself in the coherent interference of WF and UR. A pure UR takes place only in the relatively ultra-short wave range where the wave diffraction can be neglected. The power of coherent UR emitted by a bunch of $N$ particles is proportional to $N^4$ [4]. At incoherent emission the UR power is proportional to $N^3$ [5].

In the present paper we will consider UR spectral–angular characteristics for relativistic charged particles interacting with transverse components of nonsynchronous spatial harmonics of the self-wakefields excited in a periodic structure. Also we will analyze conditions when incoherent UR power can exceed beam loss power needed for exciting wakefields.

## 2. RADIATION BY SINGLE-PARTICLE

We will consider a vacuum corrugated waveguide with metallic surface as a periodic structure. Let a particle with the ultrarelativistic longitudinal velocity $v_0$, the charge $e$, and the mass $m$ moves through the structure with the period, $D$. The UR power emitted by the particle in the spectral region where wave diffraction can be neglected ($\omega_{pe} < \omega$), is given in the dipole approximation in the form [4]

$$P_{UR} = \frac{e^2}{16\pi c} \int_0^{2\pi} d\varphi \int_0^\pi d\theta \sin\theta \int_{\omega_{pe}}^{\gamma c/r_0} \omega^2 d\omega \sum_{p \ne 0} \left\{ \frac{|\alpha_x^{(p)}|^2}{p^2} [1 - \sin^2\theta \cos^2\varphi R(\omega,\theta,p)] \right.$$
$$\left. + \frac{|\alpha_y^{(p)}|^2}{p^2} [1 - \sin^2\theta \sin^2\varphi R(\omega,\theta,p)] - \text{Re}\left(\frac{\alpha_x^{(p)} \alpha_y^{(p)}}{p^2}\right) \sin 2\varphi \sin^2\theta R(\omega,\theta,p) \right\}$$
$$\times \left\{ \delta[\omega(\beta_0 \cos\theta - 1) - p\Omega] + \delta[\omega(\beta_0 \cos\theta + 1) - p\Omega] \right\}$$

(1)

Here $R(\omega,\theta,p) \equiv \left(1 - \frac{\omega}{\Omega p}\beta_0 \cos\theta\right)^2 - \beta_0^2 \left(\frac{\omega}{\Omega p}\right)^2$, $\Omega = 2\pi v_0/D$, $\theta$ is the angle between wave vector $\boldsymbol{k}$ and the longitudinal axis OZ, $\omega$ is the frequency, $\omega_{pe}$ is the electron plasma frequency of the waveguide metal, $\varphi$ is the angle between the axis OX and XOY - plane projection of $\boldsymbol{k}$, $\boldsymbol{\alpha}^{(p)} \equiv \frac{2e^2 \boldsymbol{u}_\perp^{(p)}}{mc\gamma\Omega}$ is a small parameter $|\alpha^{(p)}| \ll 1$, $\boldsymbol{u}_\perp^{(p)}$ is the transverse component of the $p^{th}$ harmonic of the wake function defined as $\boldsymbol{u}_\perp^{(p)} = \boldsymbol{w}_\perp^{(p)} + \boldsymbol{w}_\perp^{(-p)*}$, where $\boldsymbol{w}^{(p)}$ is

$$\boldsymbol{w}^{(p)} \equiv \frac{Dv_0}{4c^2 V_{cell}} \sum_{n=0}^\infty \sum_{\lambda_j'} \frac{g_{z,\lambda_j}^{(n)*}}{\left|v_0 - \frac{d\omega_\lambda}{dh}\right|_{\lambda=\lambda_j}} \left[ \boldsymbol{g}_{z,\lambda_j}^{(n+p)} - i\frac{v_0}{\omega_{\lambda_j}} \nabla_\perp g_{z,\lambda_j}^{(n+p)} - \frac{\Omega p}{\omega_{\lambda_j}} \boldsymbol{g}_{\perp,\lambda_j}^{(n+p)} \right]$$

$\boldsymbol{g}_\lambda^{(n)} \equiv \boldsymbol{g}_\lambda^{(n)}(\boldsymbol{r}_{0,\perp})$ is the $n^{th}$ spatial harmonic of the vector potential eigenfunction, $h$ is the propagation constant, $c$ is the velocity of light, $\gamma$ is Lorentz's factor, $\omega_{\lambda_j}$ set of the eigenfrequencies satisfying the well known resonant radiation conditions of a charged particle moving through a periodic structure, $\omega_\lambda - h(\omega_\lambda)v_0 = n\Omega$.

Integrating Eq.(1) over the frequency $\omega$ and angle $\varphi$ we find the resonant UR frequencies

$$\omega^{(p)} = \frac{|p|\Omega}{1 - \beta_0 \cos\theta}, \qquad (2)$$

and the angle distribution of UR power

$$P_{UR} = \frac{e^2 \Omega^2}{4c}\left\{\sum_{p=p_{min}}^{p=p_0}|\alpha^{(p)}|^2 \int_0^{\theta_p} d\theta \frac{\sin\theta}{|1-\beta_0\cos\theta|^3}\left(1-\frac{(1-\beta_0^2)\sin^2\theta}{2(1-\beta_0\cos\theta)^2}\right)\right.$$
$$\left. + \sum_{p=p_0}^{p<<p_{lim}}|\alpha^{(p)}|^2 \int_0^{\pi} d\theta \frac{\sin\theta}{|1-\beta_0\cos\theta|^3}\left(1-\frac{(1-\beta_0^2)\sin^2\theta}{2(1-\beta_0\cos\theta)^2}\right)\right\} \qquad (3)$$

Here,

$$\theta_p = \arccos\left[\frac{1}{\beta_0}\left(1-\frac{p\Omega}{\omega_{pe}}\right)\right], \quad p_0 \text{ is integer of } \left[\frac{\omega_{pe}}{\Omega}(1+\beta_0)\right],$$

$$p_{min} = \begin{cases} 1, & \text{if } (1-\beta_0)\omega_{pe}/\Omega \leq 1 \\ \text{integer of }[(1-\beta_0)\omega_{pe}/\Omega], & \text{if } (1-\beta_0)\omega_{pe}/\Omega > 1 \end{cases}$$

The number of harmonics in the sum is defined by the dipole limit resulting in $p << p_{lim} = 2\pi\gamma/\max\{\alpha^{(p)}\}$.

Integrating Eq.(1) over $\theta$ and $\varphi$, the spectrum distribution of the UR power is obtained in the form

$$P_{UR} = \frac{e^2}{4c}\left\{\sum_{p=p_{min}}^{p=p_0-1}\frac{|\alpha^{(p)}|^2}{p}\int_{\omega_{pe}}^{\frac{p\Omega}{1-\beta}} d\omega \frac{\omega}{\beta_0}\left\{1-\frac{1}{2}\left(\frac{\omega}{\gamma p\Omega}\right)^2\left[1-\frac{1}{\beta_0^2}\left(1-\frac{p\Omega}{\omega}\right)^2\right]\right\}\right.$$
$$\left.+\sum_{p=p_0+1}^{p<<p_{lim}}\frac{|\alpha^{(p)}|^2}{p}\int_{\frac{p\Omega}{1+\beta}}^{\frac{p\Omega}{1-\beta}} d\omega \frac{\omega}{\beta_0}\left\{1-\frac{1}{2}\left(\frac{\omega}{\gamma p\Omega}\right)^2\left[1-\frac{1}{\beta_0^2}\left(1-\frac{p\Omega}{\omega}\right)^2\right]\right\}\right\} \qquad (4)$$

It the case of high energy particles satisfying the condition $\omega_{pe} << 2\Omega\gamma^2$, Eqs. (3) and (4) results in

$$P_{UR} \approx \frac{4e^6}{3m^2c^3}\gamma^2 \sum_{p=1}^{p<<p_{lim}}|u_\perp^{(p)}|^2. \qquad (5)$$

## 3. RADIATION BY BUNCH

As follows from Eq.(5), if there is a bunch of $N$ electrons with the longitudinal and transverse dimensions $\sigma_z$ and $\sigma_\perp$ satisfying the conditions $\sigma_z << D/(2q\gamma^2)$ and $\sigma_\perp << D/(2q\gamma)$, respectively, then the total UR power is coherent and proportional to $N^4$

$$P_{UR} = \frac{4e^6 N^4}{3m^2c^3}\gamma^2 \sum_{p=1}^{q}|u_\perp(p)|^2. \qquad (6)$$

However the dimensions $\sigma$ of bunches accelerated in the high energy rf linacs satisfy the relation $D/\gamma^2 << \sigma$ as well as $\omega_{pe} << 2\Omega\gamma^2$. So, it is of interest to consider features of incoherent hard radiation emitted by existing beams. The incoherent UR power may be written as [5]

$$P_{UR} = \frac{4e^6}{3m^2c^3}N^3 \sum_{p=1}^{p<<p_{lim}}\int_{-\infty}^{\infty}d\tau\iint_{S_\perp} d^2\boldsymbol{r}_\perp f_b(\boldsymbol{r}_\perp,\tau)\gamma(\boldsymbol{r}_\perp,\tau)^2|\boldsymbol{u}_\perp^{(p)}(\boldsymbol{r}_\perp,\tau)|^2, \qquad (7)$$

while the WF power loss is given by

$$P_{WF} = v_0(eN)^2 \int_{-\infty}^{\infty}d\tau \iint_{S_\perp} d\boldsymbol{r}_\perp f_b(\boldsymbol{r}_\perp,\tau)u_z^{(0)}(\boldsymbol{r}_\perp,\tau) \qquad (8)$$

where $f_b(\boldsymbol{r}_\perp,\tau)/v_0$ is the normalized function of charge-density distribution, $S_\perp$ is the cross-section of the periodic structure, $\tau = t - z/v_0$ is time when the particle crosses the plane z=0, the $p^{th}$ harmonic of the wake function is defined as

$$\boldsymbol{u}^{(p)}(\boldsymbol{r}_\perp,\tau) \equiv -\frac{1}{e^2 N}\int_0^{\infty}d\tau'\iint_{S_\perp} d^2\boldsymbol{r}'_{0\perp}f_b(\boldsymbol{r}'_{0\perp},\tau-\tau')\boldsymbol{F}^{(p)}(\boldsymbol{r}_\perp,\boldsymbol{r}'_{0\perp},\tau') \qquad (9)$$

$\boldsymbol{F}^{(p)}(\boldsymbol{r}_\perp,\boldsymbol{r}_{0\perp},\tau)$ is the $p$th spatial harmonic of the force, produced by the point charge $eN$ (with the transverse coordinate $\boldsymbol{r}_{0\perp}$) acting on the test charge $e$ (with the transverse coordinate $\boldsymbol{r}_\perp$) moving at the distance $v_0\tau$ after the point charge.

For analytical calculations it is convenient to consider a monochromatic infinitesimally thin uniform bunch of the length $l_b$ moving in a weakly corrugated circular waveguide of the radius

$$b(z) = b_0[1+\varepsilon(z)] = b_0\left[1+\sum_{p=-\infty}^{\infty}\varepsilon_p \exp\left(i\frac{2\pi p}{D}z\right)\right]. \qquad (10)$$

Here $\varepsilon(z) << 1$ is the relative depth of corrugations, $b_0$ is the average radius of the waveguide.

The UR power lost by the bunch moving at the distance $r_b$ from the waveguide axis is given by [5]

$$P_{UR} = \frac{4e^6}{3m^2c^3}\gamma^2 N^3 \left(\frac{4\pi}{b_0 D}\right)^2 \sum_{p=1}^{\infty}p^2|2\varepsilon_p|^2 \sum_{m=0}^{\infty}\sum_{n=0}^{\infty}\frac{(r_b/b_0)^{m+n}}{(1+\delta_{0,m})(1+\delta_{0,n})}$$
$$\times \sum_{s=1}^{\infty}\sum_{q=1}^{\infty}\{A_{m,s}A_{n,q}I(\omega_{m,s,p},\omega_{n,q,p})+B_{m,s}B_{n,q}I(\omega'_{m,s,p},\omega'_{n,q,p}) \qquad (11)$$
$$-A_{m,s}B_{n,q}I(\omega_{m,s,p},\omega'_{n,q,p})-B_{m,s}A_{n,q}I(\omega'_{m,s,p},\omega_{n,q,p})\}$$

where $\delta_{0,m}$ is Chronicler's symbol,

$$I(\omega_1,\omega_2) \equiv \frac{c^2}{\omega_1\omega_2 l_b^2}\left[1-\frac{\sin(\omega_1 l_b/c)}{\omega_1 l_b/c}-\frac{\sin(\omega_2 l_b/c)}{\omega_2 l_b/c}+\frac{\sin((\omega_1-\omega_2)l_b/c)}{(\omega_1-\omega_2)l_b/c}\right],$$

$$A_{m,s}(r/b_0) \equiv \frac{J'_m(\mu_{m,s} r/b_0)}{J'_m(\mu_{m,s})}, \quad B_{m,s}(r/b_0) \equiv \frac{b_0}{r}\frac{m^2}{m^2-\mu'^2_{m,s}}\frac{J_m(\mu'_{m,s} r/b_0)}{J_m(\mu'_{m,s})},$$

$\mu_{m,s}$ and $\mu'_{m,s}$ are the zeros of Bessel functions $J_m(\mu_{m,s}) = 0$ and $J'_m(\mu'_{m,s}) = 0$, respectively,

$$\omega_{m,s,p} = \frac{\pi pc}{D}\left[1+\left(\frac{D\mu_{m,s}}{2\pi p b_0}\right)^2\right] \quad \omega'_{m,s,p} = \frac{\pi pc}{D}\left[1+\left(\frac{D\mu'_{m,s}}{2\pi p b_0}\right)^2\right]$$

are the frequencies of resonant WF modes. Therewith, the WF power loss of the bunch is [5]

$$P_{WF} = \frac{2\pi(eN)^2}{D}\sum_{p=1}^{\infty}p|2\varepsilon_p|^2 \sum_{m=0}^{\infty}\left(\frac{r_b}{b_0}\right)^{2m}\frac{1}{(1+\delta_{0,m})} \qquad (12)$$
$$\times\sum_{s=1}^{\infty}\left\{\omega_{m,s,p}\left(\frac{\sin(\omega_{m,s,p}l_b/2c)}{\omega_{m,s,p}l_b/2c}\right)^2 - \frac{m^2\omega'_{m,s,p}}{m^2-\mu'^2_{m,s}}\left(\frac{\sin(\omega'_{m,s,p}l_b/2c)}{\omega'_{m,s,p}l_b/2c}\right)^2\right\}$$

Let us consider a sinus-type corrugated waveguide of the radius $b(z) = b_0[1+2\varepsilon_1\sin(2\pi z/D)]$, with $\varepsilon_1 = 0.05$. Let an electron bunch, with the typical for SLAC Linac parameters [7], $l_b = 500$ μm, $N = 4\times 10^{10}$, and $\gamma = 10^5$ moves in the "sub-millimeter" structure with $b_0 = D = 0.3$ mm. The bunch distance from the axis is chosen equal $r_b = 0.9b_0$ to estimate the maximal values of the UR power.

The distribution of the synchronous harmonic of longitudinal wake function $u_z^{(0)}$ and the first harmonic transverse wake function $u_r^{(+1)}$ along the bunch are represented in Figs.1 and 2, respectively. In calculating the first 120 resonant WF modes are taken account.

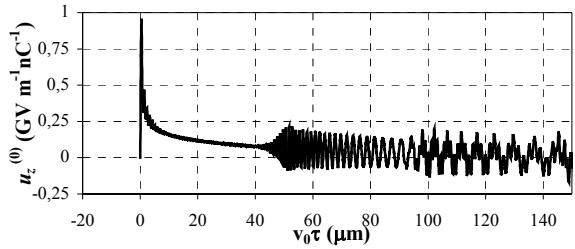

***Fig. 1.** The synchronous harmonic of longitudinal wake function.*

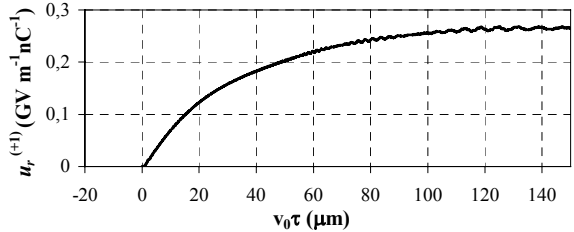

***Fig. 2.** The nonsynchronous harmonic of transverse wake function.*

As shown in Fig.1, the bunch head, up to 50 μm, basically excites WF while UR is predominantly emitted by the next part after head of the bunch (see Fig.2).

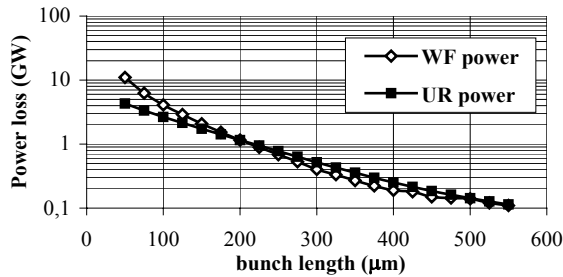

***Fig. 3**. The UR and WF power vs. the bunch length.*

Eqs. (5)-(7) shows that in the range of ultra-sort wavelength light the UR power grows as square of particle energy. Therefore, we can expect, that for very high energy particles the UR power can exceed the WF power which, as well known, is independent on the particle energy. Figs. 1, 2 indicate a possible increase of the relative fraction of the UR power as the bunch length grows. This assumption is confirmed by Fig. 3. As seen, the incoherent UR power of 50 GeV beam reaches the WF power at the bunch length 200 μm.

The incoherent UR power and WF power for the bunch of 500 μm length versus electron energy are represented in Fig. 4. As seen from this figure the UR predominates at energies above of 50 GeV.

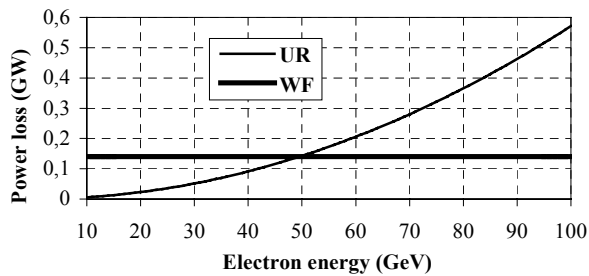

***Fig. 4**. The incoherent UR and WF power vs. the beam energy.*

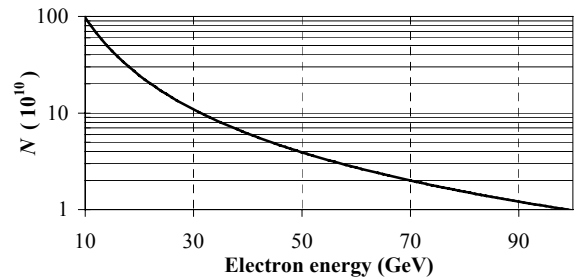

***Fig. 5.** The number of bunch electrons vs. the electron energy when the UR power equals the WF power.*

In the Fig. 5 we represent the dependence of a number of electrons in the bunch of length 500 μm on their energy in the case when the incoherent UR power equals WF power. This dependence determines the limit, above of which the incoherent UR power exceeds the power of the wakefield excitation.

## 4. CONCLUSION

The spectral–angular characteristics of the undulator-type radiation of an ultrarelativistic charged particle undulating in nonsynchronous spatial harmonics of the self-wakefields in the periodic structure have been obtained. Using the model of a weakly corrugated circular waveguide excited by an infinitesimally thin uniform electron bunch, it has been shown numerically that whereas the beam power losses fall as the bunch length grows, the relative fraction of the UR power increases. The conditions when the incoherent UR power becomes comparable with the WF power have been found. These conditions are acceptable for possible experimental verification at SLAC Linac [7].